\documentclass[jkps,twocolumn,fleqn,showkeys]{revtex4-2}
\usepackage{amssymb}
\usepackage{amsmath}
\usepackage{graphicx}
\usepackage{bm,multirow}
\usepackage[]{color}

\begin{document}
%\setcounter{page}{0}
%\preprint{}
\title{Structural Investigation of BaIrO$_3$ by Neutron Diffraction}


\author{Bin Chang, Jinwon Jeong, Han-Jin Noh}
\email{ffnhj@jnu.ac.kr}
\affiliation{Department of Physics, Chonnam National University, Gwangju 61186, Korea}
\author{Seongsu Lee}
\affiliation{Neutron Science Division, Korea Atomic Energy Research Institute, Daejeon 34057, Korea}

%\thanks{}

\date{\today}

\begin{abstract}
We report a  temperature-dependent neutron diffraction (ND) study on polycrystalline monoclinic BaIrO$_3$ which is famous for charge density wave (CDW) and weak ferromagnetic phase transitions at T$_C$$\sim$180 K simultaneously.
A Rietveld analysis on the ND patterns reveals that even though there is no symmetry breaking in crystal structure, a noticeable change in the four kinds of IrO$_{6}$ octahedra is isolated as the temperature approaches to T$_C$.
Based on the structure analysis results, we calculated the $d$-orbital energy level splittings by crystal electric field for each type of the IrO$_6$ octahedra.
By taking into account the strong spin-orbit coupling in Ir 5$d$ orbitals and the lattice distortions obtained from the ND analysis, we propose an electronic configuration model to understand the phase transition of the system, where an effective $J_{\rm eff, 1/2}$ Mott insulating phase and a charge gap phase induced by bonding states between the $J_{\rm eff,1/2}$ states compete each other.
\end{abstract}

%\pacs{71.45.Lr, 61.05.-a, 65.40.De, 79.60.-i}
% 71.45.Lr charge density wave systems
% 61.05.-a techniques for structure determination
% 65.40.De thermal expansion of crystalline solids
% 79.60.-i photoemission and photoelectron spectra
%\keywords{}

\maketitle

%\section{Introduction}
Transition metal oxides (TMO's) have been an ideal platform for investigating strong electron correlations in condensed matter systems.\cite{Cox}
At the early stage of the research, much attention was focused on 3$d$ TMO's, where Coulomb interactions and exchange energies in 3$d$ electrons are the key sources for the strong correlations.
The prototypical phenomena in 3$d$ TMO's such as Mott insulating phase, charge transfer insulating phase, and metal-insulator transition become understandable only after introducing those correlations. 
However, little progress had made on 4$d$ and 5$d$ TMO's until the exact role of spin-orbit interactions (SOI) in band electrons was identified.\cite{Kim1, Kim2}
The discovery of the effective total angular momenta in the Ir 5$d$ orbitals of Sr$_2$IrO$_4$ ignited 4$d$ and 5$d$ TMO's researches in view of the SOI physics.
 
Quasi-one-dimensional monoclinic BaIrO$_3$ is an another prototypical example to investigate the SOI physics and the correlation effects in 5$d$ TMO's since it has strong SOI in Ir 5$d$ states and low dimensionality in the crystal structure.
As is expected, various intriguing physical properties have been measured in this compound.
In particular, it is known that weak ferromagnetism and charge density waves (CDW) coexist below $\sim$180 K\cite{Cao1}, but its exact ground state has not been fully understood.
In spite of the long history of research on this compound, there remain a few controversal issues that are not clearly resolved yet.
For examples, the exact charge modulation vectors and the driving force of CDW, the ordering pattern of the magnetic moments, and the exact coupling mechanism between ferromagnetic order and CDW remain unclear or do not reach on a consensus.

The CDW phase was proposed based on several experimental results. 
A discontinuous increase in the resistivity curve, an abrupt feature in the non-linear conductivity showing negative differential resistivity, and a gap formation at $\sim$1200 cm$^{-1}$ in the optical spectra were provided in the previous study.\cite{Cao1}
However, the resistivity vs. temperature curve shows insulating phases both above and below T$_C$, which is contrastive to a conventional CDW phase where a metallic system gets a lowered energy state by an energy gap opening induced by a charge modulation.
Further, superstructure peaks in diffraction patterns or charge disproportionation in photoemission spectra have not been observed so far.
Due to these controversal experimental results, the CDW phase was questioned in the I-V measurement study, and alternative models based on phase separation or on electron liquid crystal were proposed in order to explain the giant nonlinear conductivity.\cite{Nakano}

The ferromagnetic phase of the compound below the transition temperature (T$_{C}$ $\sim$180 K) has been established in 1990's by the Faraday method since the correct crystal structure analysis was performed through the several preliminary studies.\cite{Siegrist, Sarkozy, Lindsay}
However, the weak ferromagnetism and the magnetic moment reduction from the paramagnetic to ferromagnetic transition has not been understood.
The electronic structure calculated by the local spin density approximation (LSDA) method indicates that the system has itinerant electrons with small spin polarization in Ir 5$d$ bands, which is incompatible with the canted spin 1/2 model proposed by Lindsay {\it et al}.\cite{Maiti1, Lindsay}
An x-ray magnetic circular dichroism (XMCD) study revealed that the large orbital magnetic moments are dominant in the magnetism of this system.\cite{Laguna-Marco}
Also, there is a report that unconventional critical behavior was observed in the weak ferromagnetic BaIrO$_3$.\cite{Kida}
Recently, Terasaki {\it et al.} re-investigated the origin of the phase transition at $\sim$180 K by measuring resistivity, thermopower, magnetization and synchrotron X-ray diffraction of Ru-doped BaIrO$_3$, and concluded that it is a transition from a Mott insulator phase to a charge ordered insulator phase. \cite{Terasaki}

In this study, we investigate the relation between the electronic configuration of the ground states and the structural change of IrO$_6$ trimers in monoclinic BaIrO$_3$ by temperature-dependent neutron diffraction (ND).
Although a temperature-dependent synchrotron XRD study was reported\cite{Terasaki}, ND study is inevitable to determine the exact positions of Ir and O atoms due to the high cross section of neutrons to oxygens.\cite{NeutronCross}
Surprisingly, to the best of our knowledge, there has been no temperature-dependent ND study so far even though ND data at one temperature point in one structural phase were reported.\cite{Cheng2} 
Our Rietveld refinements on the ND patterns revealed that there was no symmetry breaking across the phase boundary in crystal structure, but that a noticeable change was successfully isolated in the IrO$_6$ octahedra across T$_C$, which can lead to a change of the electronic configuration in the band structure of quasi-one-dimensional monoclinic BaIrO$_3$. 
By taking into account the Ir site-selective lattice distortion, the spin-orbit coupling of Ir 5$d$ orbitals, and the quasi-one-dimensional structure altogether, we propose a model for the electronic configuration where a $J_{\rm eff, 1/2}$ Mott insulating phase and a charge gap phase induced by the bonding states of the effective $J_{\rm eff, 1/2}$ states compete each other.

The polycrystalline samples of monoclinic BaIrO$_3$ were synthesized by a standard solid state reaction method.
Stoichiometric amounts of high purity ($\geq$ 99.99 \%) BaCO$_3$ and Ir metal powders were weighed and mixed with a pestle and mortar.
The mixtures were fired three times at temperature of 1000 $^{\circ}$C for two days.
The neutron powder diffraction experiments were performed on the high resolution powder diffraction (HRPD) neutron beam line at the HANARO research reactor, Korea.
The wave length of the monochromatized neutrons, which is achieved by single crystal Ge 331 lattice planes, is 1.8345 \AA.

\begin{figure}
\includegraphics[width=8.0 cm]{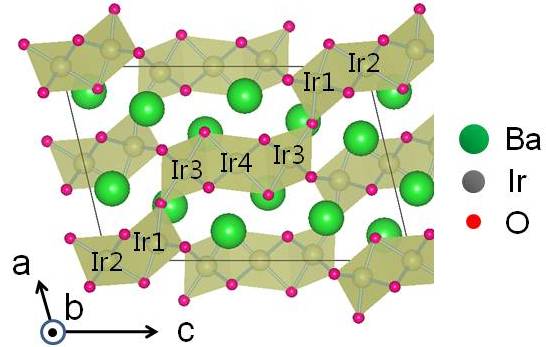}
\caption{\label{fig1}
Crystal structure of 9R-type monoclinic (C$_{2/m}$) BaIrO$_3$. Ir1 - Ir4 denote the distinct Ir sites in a unit cell. Solid parallelogram denotes the unit cell.
}
\end{figure}

Figure~\ref{fig1} shows the crystal structure of the monoclinic BaIrO$_3$ belonging to space group C$_{2/m}$.
This structure is isostructural with so-called 9R-type BaRuO$_3$ except for the monoclinic distortion\cite{Cheng2}, which was firstly determined by Siegrist and Chamberland.\cite{Siegrist}
Among the polytypes of BaIrO$_3$, this 9R-type barium iridate has the most one-dimensional-like structure.\cite{Cheng1, Cheng2}
Each Ir atom is located inside an octahedron, and three IrO$_6$ octahedra form a so-called Ir$_3$O$_{12}$ trimer by face-sharing.
The trimers are linked with each other by sharing one corner of oxygen atom as shown in Fig.~\ref{fig1}.
Crystallographically, there are four independent Ir sites (Ir1 - Ir4), six oxygen sites (O1 - O6), and two kinds of Ir trimers in a unit cell.
Note that Ir1 and Ir3 are different from Ir2 and Ir4 in that odd number Ir sites are more sensitive to trimer distortion.
This is an important clue to understand the temperature dependent structure study described below.

\begin{figure}
\includegraphics[width=8.5 cm]{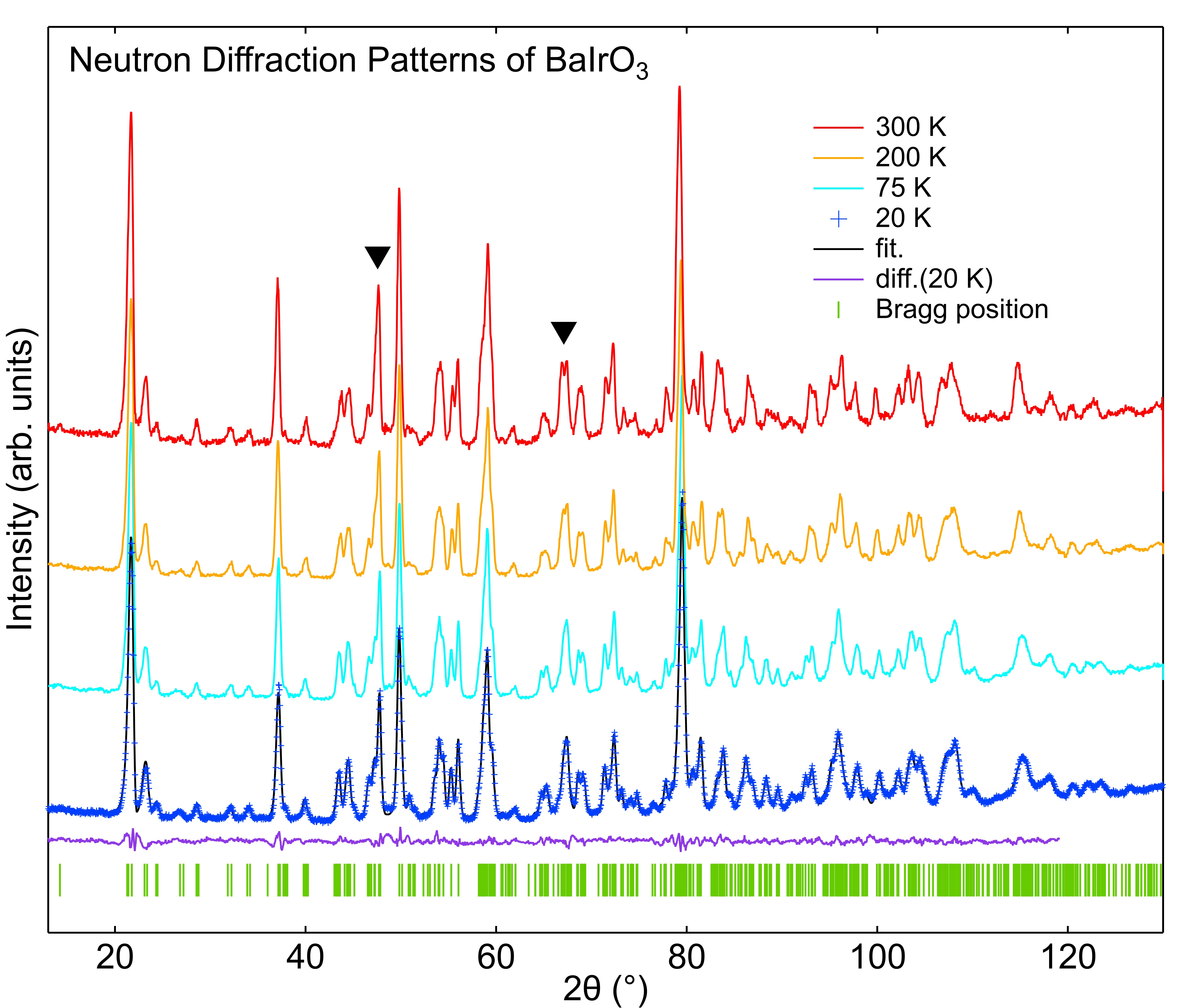}
\caption{\label{fig2}
Neutron diffraction patterns of monoclinic BaIrO$_3$ at 20, 75, 200, and 300 K. The Rietveld refinement result is shown only for the 20 K data for clarity. The peaks denoted with a black triangle ($\blacktriangledown$) are the examples for the temperature-dependent structural evolution.
}
\end{figure}

The neutron diffraction patterns at 300, 200, 75, and 20 K are presented in Fig.~\ref{fig2}.
With these ND data, we performed Rietveld refinements to investigate temperature-dependent structural changes using Fullprof 2000.\cite{Roisnel}
For simplicity, we present the refinement results only for the 20 K data in Fig.~\ref{fig2}.
The goodness of fit for the other temperature data is of the same order.
For the estimation, weighted profile R-factor R$_{wp}$ and chi-square value $\chi^{2}$ are listed in Table \ref{table1}.
In the patterns, we can clearly see evolutions in the peak position and peak weight.
For example, the peak ($ \blacktriangledown$) at 2$\theta \simeq$70$^{\circ}$ is split as the temperature decreases.
However, those split peaks correspond not to superstructure peaks but to internal Wyckoff position changes.
All our ND patterns are successfully reproduced by adjusting only the cell parameters and atom positions under monoclinic C$_{2/m}$ structure.
The detailed refinement results are available in Table \ref{table1}.

\begin{table*}
\caption {\label{table1} The refined lattice parameters of monoclinic (C$_{2/m}$) 9R-BaIrO$_3$. The number in parenthesis is the standard deviation of the last digit.}
\begin{tabular}{c | c c c c }
\hline
T   &  20 K  & 75 K & 200 K & 300 K \\
\hline
a (\AA) & 9.9915(8) & 9.9929(7) & 9.9995(7) & 10.0085(8) \\
b (\AA) & 5.7343(4) & 5.7356(4) & 5.7460(3) & 5.7541(4) \\
c (\AA) & 15.238(1) & 15.236(1) & 15.1955(9) & 15.180(1) \\
$\beta (^\circ)$ & 103.385(4) & 103.399(3) & 103.318(3) & 103.258(4) \\
\hline
Ba1 & 0.776(1),0,0.2414(8) & 0.773(1),0,0.2486(7) & 0.760(2),0,0.245(1) & 0.756(2),0,0.242(1) \\
Ba2 & 0.373(1),0,0.0750(7) & 0.374(1),0,0.0794(8) & 0.379(1),0,0.0786(7) & 0.376(1),0,0.077(1) \\
Ba3 & 0.162(1),0,0.4294(7) & 0.1564(9), 0, 0.4249(7) & 0.157(1), 0, 0.4290(6) & 0.161(1),0,0.4281(9) \\
\hline
Ir1 & 0.0880(7),0,0.1733(4) & 0.0863(6), 0, 0.1722(3) & 0.0879(7), 0, 0.1713(4) & 0.0861(8),0,0.1715(4) \\
Ir2 & 0,0,0 & 0,0,0 & 0,0,0 & 0,0,0 \\
Ir3 & 0.4706(8),0,0.3207(5) & 0.4701(7), 0, 0.3189(5) & 0.4635(8), 0, 0.3220(5) & 0.463(1),0,0.3236(6) \\
Ir4 & 0.5, 0, 0.5 & 0.5, 0, 0.5 & 0.5, 0, 0.5 & 0.5, 0, 0.5 \\
\hline
O1 & 0.301(1),0,0.2266(7) & 0.301(1),0,0.2257(6) & 0.304(1),0,0.2270(9) & 0.301(2),0,0.230(1) \\
O2 & 0.0578(6),0.232(1),0.2623(5) & 0.0580(6),0.240(1),0.2616(4) & 0.0540(6),0.236(1),0.2602(4) & 0.0506(8),0.232(2),0.2590(5) \\
O3 & 0.888(1),0,0.0986(6)) & 0.891(1),0,0.09832(9) & 0.891(1),0,0.0953(9) & 0.891(1),0,0.0936(9) \\
O4 & 0.1225(9),0.231(2),0.0848(5) & 0.1189(8),0.237(1),0.0815(7) & 0.1187(8),0.230(2),0.0813(6) & 0.120(1),0.228(2),0.084(1) \\
O5 & 0.4046(7),0.224(1),0.4041(4) & 0.4074(6),0.231(1),0.4000(3) & 0.4068(6),0.230(2),0.4005(3) & 0.4049(8),0.229(2),0.4024(4) \\
O6 & 0.647(1),0,0.4313(8) & 0.637(2),0,0.4233(9) & 0.642(2),0,0.425(1) & 0.643(2),0,0.425(1) \\
\hline
R$_{wp}$,  $\chi{^2}$   &  7.98, 5.974  & 8.06, 5.953 & 8.19, 5.774 & 10.3, 3.483 \\
\hline
\end{tabular}

\end{table*}

\begin{figure}
\includegraphics[width=8.5 cm]{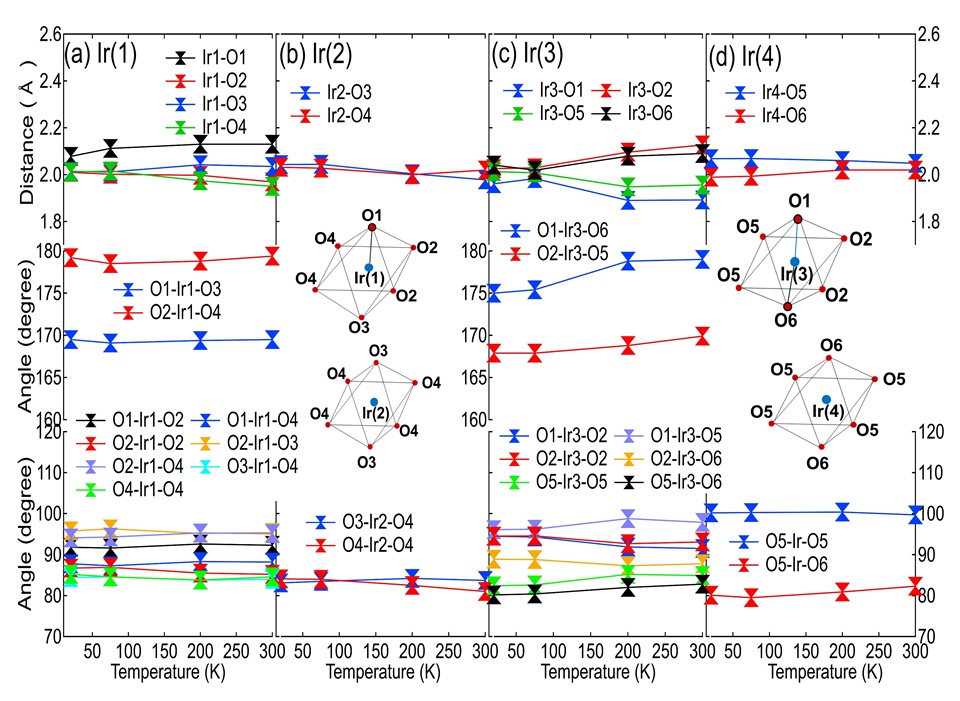}
\caption{\label{fig3}
Bonding distances between Ir and O atoms, and O-Ir-O angles in the IrO$_6$ octahedra for (a) Ir1, (b) Ir2, (c) Ir3, and (d) Ir4 as a function of temperature extracted from the Rietveld analysis of the ND patterns. The cartoons for IrO$_6$ octahedra show what kinds of oxygen sites surround the iridium site. 
}
\end{figure}

\begin{figure}
\includegraphics[width=8.5 cm]{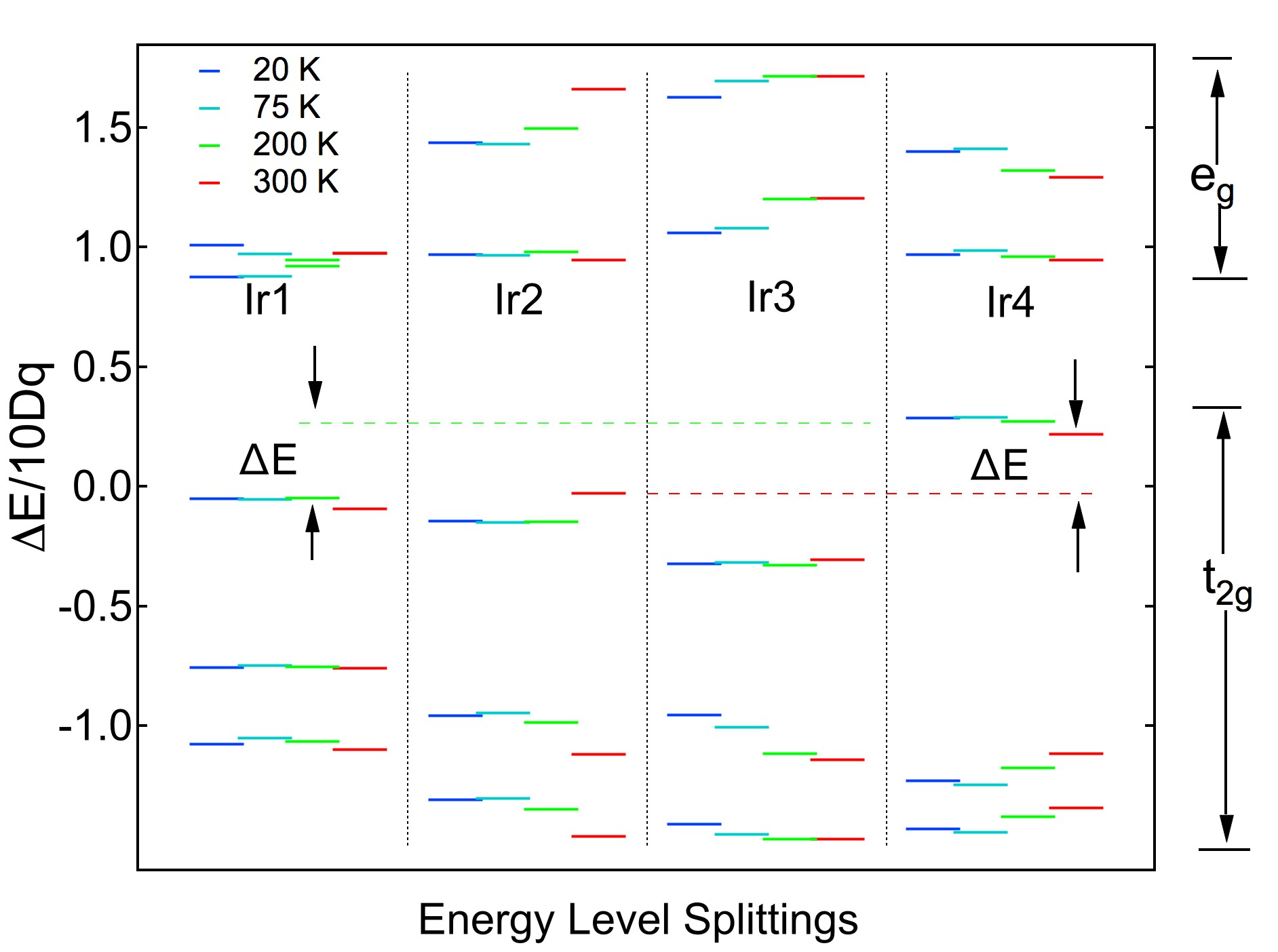}
\caption{\label{fig4}
Ir 5$d$ level splittings induced by the crystal electric field on each type of IrO$_6$ octahedron.
Noticeable level changes in Ir1, Ir2 and Ir4 denoted by two arrows are isolated in the calculation.
}
\end{figure}

In Fig.~\ref{fig3}, we present all the bonding distances between Ir and O atoms, and all the angles of O-Ir-O in monoclinic BaIrO$_3$ as a function of temperature, which were extracted from the refinement of the ND patterns.
The correlation between the distortion of IrO$_6$ octahedra and the CDW/magnetic phase transition can be seen from several distinctive bonding lengths and bonding angles.
For examples, the length of Ir1-O1, Ir2-O3, all Ir3-O's, and Ir4-O6 are relatively sensitive to the temperature change.
In bonding angle wise,  the Ir3 octahedra are more sensitive to temperature change than the others.
As a whole, the response of the IrO$_6$ octahedra to the temperature change is not of the first order type and very diverse in their bonding angles and lengths, which makes it hard to find an apparent connection between the structure distortion and the phase transition.
In order to circumvent the difficulty, we checked the distortion effect of the IrO$_6$ octahedra by calculating the energy level splittings of Ir 5$d$ orbitals induced by the crystal electric field (CEF) of the surrounding oxygen anions for each type of Ir cations.
In the calculation, we assumed atomic 5$d$ orbitals and O$^{2-}$ states, and employed the atomic positions obtained from the ND refinement.
The calculated energy levels as a function of temperature for each Ir cation is displayed in Fig.~\ref{fig4}, where a noticeable change of the energy difference, $\Delta E$, between the $J_{\rm eff, 1/2}$ states in Ir1-Ir2-Ir1 (trimer A) and Ir3-Ir4-Ir3 (trimer B) is revealed as denoted by the black arrows. 
The energy difference, $\Delta E$, at T=300 K (red lines) gets $\sim$20\% larger at 200 K, and the increasing tendency persists to 20 K.
Note that the zero point in energy reference is the center of the Ir 5$d$ orbitals, so the actual energy difference can be very different from this calculations, but the relative tendency remains the same.    
Taking all implications from the calculations based on Ir site selective lattice distortion into account, we propose a model that can explain why the system shows a CDW-like behavior that does not have a superstructure peak and why it has such a small magnetic moment in the ferromagnetic phase as is described below.

\begin{figure}
\includegraphics[width=8.5 cm]{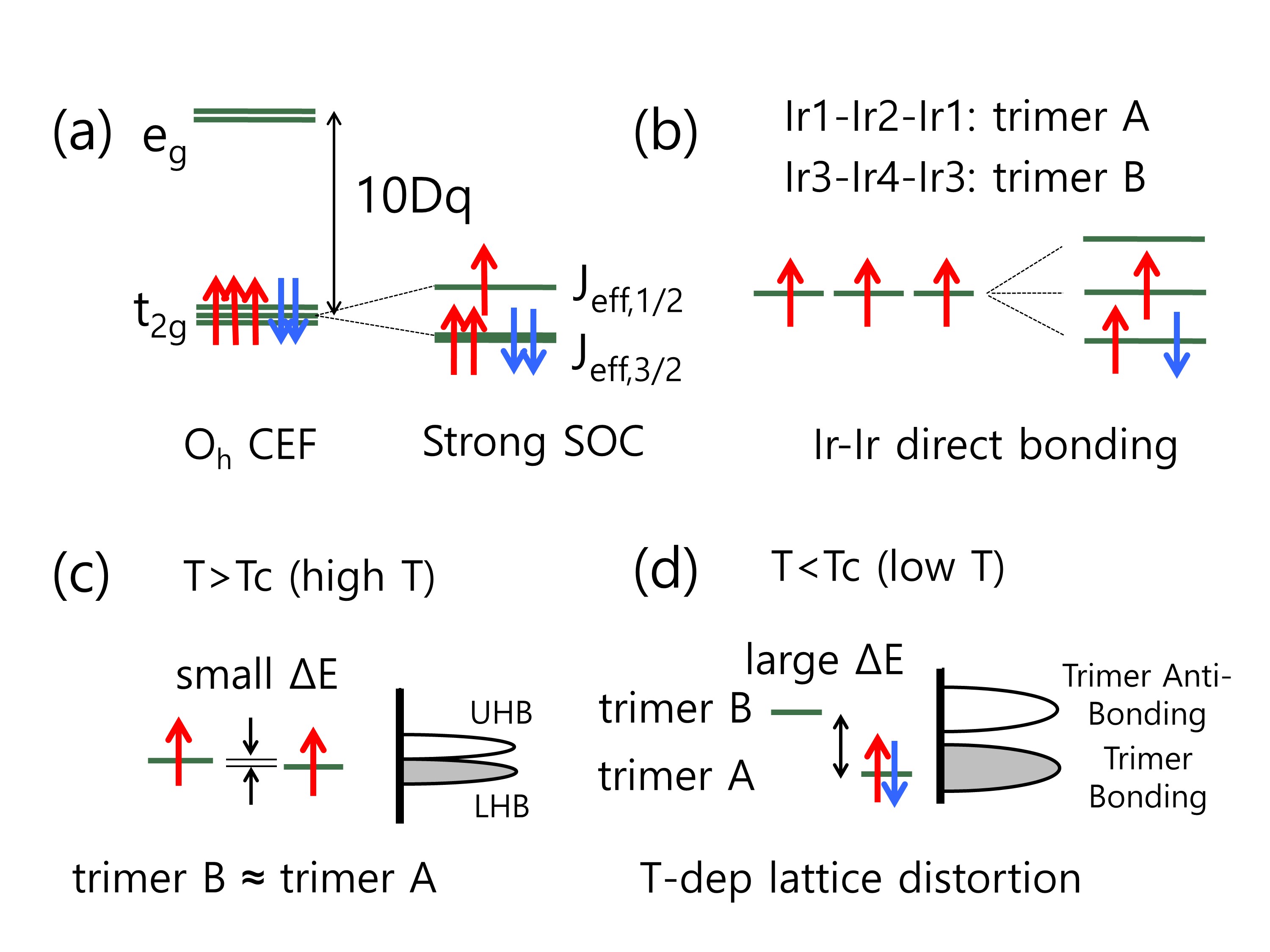}
\caption{\label{fig5}
Electronic configurations of monoclinic BaIrO$_3$.
}
\end{figure}

The Ir cations in BaIrO$_3$ are under a crystal electric field of  distorted O$_h$ symmetry, which approximately splits the Ir 5$d$ manifold into three-fold t$_{2g}$ and two-fold e$_g$ states, and five electrons are occupied the t$_{2g}$ states with a low spin configuration as shown in Fig.~\ref{fig5}(a).
In addition,  the strong SOI in 5$d$ transition metal ions stabilizes the $J_{\rm eff,1/2}$ ground state.\cite{Kim1}
Thus, each tetravalent Ir cation has a half-filled  $J_{\rm eff,1/2}$ state of which energy level is lifted a little depending on the distortion of each type of IrO$_6$ octahedra.
Spatially extended nature of 5$d$ orbitals and the face-sharing between the IrO$_6$ octahedra make a considerable overlap among the three Ir cations in a  Ir$_3$O$_{12}$ trimer, resulting in an Ir-Ir direct bonding state, a non-bonding one, and an anti-bonding one.
Thus, each trimer is regarded as a half-filled $J_{\rm eff,1/2}$ state as shown in Fig.~\ref{fig5}(b).
At a high temperature above T$_C$ (See Fig.~\ref{fig5}(c) and (d)), our ND refinements reveal that there is relatively a small energy difference between the trimers.
So, the two kinds of trimers are taken as the same molecules with a half-filled $J_{\rm eff,1/2}$ state, which forms a Mott insulating phase or a correlated metal near the transition border line due to a narrow band width resulting from the one dimensional shape of the trimers and the corner sharing linkage of the trimers.
Meanwhile, at a low temperature below T$_C$, the energy difference between the trimers A and B becomes larger, then it is energetically favorable to open a charge gap by forming a bonding state between the trimers.
This leads to a partial transfer of an electron in trimer B to A, resulting in a CDW-like behavior.
Hence, it is not a CDW but a charge disproportionation, which is observed in a few Ir compounds.\cite{Noh1}
In this case, there is no superstructure peak in diffraction patterns because the charge disproportionation period is equal to the crystal unit cell.
Our model also explains the magnetic moment reduction in the ferromagnetic phase transition.
When a trimer bonding state is fully occupied, each trimer pair has zero magnetic moment due to the singlet-like state, so in this model we need to explain why the system has a non-zero, though small, magnetic moment.
One idea is that a real ground state is composed of a superposition of the Mott state and the trimer singlet state at the low temperature.
Of course, further experimental studies are required to prove the scenario.

To summarize, we performed a temperature-dependent neutron diffraction (ND) study on polycrystalline monoclinic BaIrO$_3$.
A Rietveld analysis on the ND patterns reveals that even though there is no symmetry breaking in crystal structure, a noticeable change in IrO$_{6}$ octahedra across the transition temperature is isolated.
Based on the structure analysis results, we calculated the $d$-orbital energy level splittings by crystal electric field for each type of the IrO$_6$ octahedra.
By taking into account the strong SOI in Ir 5$d$ orbitals and the energy level shifts obtained from the calculations, we propose an electronic configuration model to understand the phase transition of the system, where $J_{\rm eff, 1/2}$ Mott states  and trimer singlet states composed of two different $J_{\rm eff, 1/2}$ states compete each other.

\acknowledgments
This work was supported by the National Research Foundation (NRF) of Korea Grant funded by the Korean Government (MEST) (No. 2011-0018808, 2019R1I1A3A0105818813).


\begin{thebibliography}{}
\bibitem{Cox} Transition Metal Oxides, P. A. Cox, Clarendon Press, Oxford (1992).
\bibitem{Kim1} B. J. Kim {\it et al}., Phys. Rev. Lett. \textbf{101}, 076402 (2008).
\bibitem{Kim2} B. J. Kim, H. Ohsumi, T. Komesu, S. Sakai, T. Morita, H. Takagi, and T. Arima, Science \textbf{323}, 1329 (2009).
\bibitem{Cao1} G. Cao et al., Solid State Commun. \textbf{113}, 657 (2000).
\bibitem{Nakano} T. Nakano and I. Terasaki, Phys. Rev. B \textbf{73}, 195106 (2006).
\bibitem{Siegrist} T. Siegrist and B. L. Chamberland, J. Less-common Metals \textbf{170}, 93 (1991).
\bibitem{Sarkozy} R. F. Sarkozy, Ph. D. Thesis, University of Connecticut (1974).
\bibitem{Lindsay} R. Lindsay, W. Strange, B. L. Chamberland, R. O. Moyer Jr., Solid State Commun. \textbf{86}, 759 (1993).
\bibitem{Maiti1} K. Maiti, Phys. Rev. B \textbf{73}, 115119 (2006).
\bibitem{Laguna-Marco} M. A. Laguna-Marco {\it et al}., Phys. Rev. Lett. \textbf{105}, 216407 (2010).
\bibitem{Kida} T. Kida, A. Senda, S. Yoshi, M. Hagiwara, T. Takeuchi, T. Nakano, and I. Terasaki, EPL \textbf{84}, 27004 (2008).
\bibitem{Terasaki} I. Terasaki {\it et al.}, Crystals \textbf{6}, 27 (2016).
\bibitem{NeutronCross} V. McLane, C. L. Dunford, and P. F. Rose, Neutron Cross Sections, Academic Press (1988).
\bibitem{Cheng2} J.-G. Cheng, J. A. Alonso, E. Suard, J.-S. Zhou, and J. B. Goodenough, J. Am. Chem. Soc. \textbf{131}, 7461 (2009).
%\bibitem{HDKim} H.-D. Kim {\it et al}., AIP conf. Proc. \textbf{879}, 477 (2007).
\bibitem{Cheng1} J.-G. Cheng, J.-S. Zhou, J. A. Alonso, J. B. Goodenough, Y. Sui, K. Matsubayashi, and Y. Uwatoko, Phys. Rev. B \textbf{80}, 104430 (2009).
\bibitem{Roisnel} T. Roisnel and J. Rodriguez-Carvajal, FullProf 2000.
%\bibitem{Ashcroft} N. W. Ashcroft and N. D. Mermin, Solid State Physics, Saunders Colledge Publishing.
%\bibitem{Kiyama} T. Kiyama, K. Yoshimura, K. Kosuge, Y. Ikeda, and Y. Bando, Phys. Rev. B \textbf{54}, R756 (1996).
\bibitem{Noh1} H.-J. Noh, E.-J. Cho, H.-D. Kim, J.-Y. Kim, C.-H. Min, B.-G. Park, S.-W. Cheong, Phys. Rev. B \textbf{76}, 233106 (2007).

\end{thebibliography}
\end{document}